# Frequency Dependence and Zero Crossover Effects in SQUID Based TDEM Measurements: Peculiarities or Facts?

R. Nagendran, Lata Bisht, Ijee Mohanty, A.V. Thanikai Arasu and Shaju K. Albert
Materials Science Group, Indira Gandhi Centre for Atomic Research,
Homi Bhabha National Institute, Kalpakkam, India

*Abstract—* In this paper we analyzed the two peculiarities frequently observed in SQUID magnetometer applications in TDEM based geophysical exploration such as the decay of the ground response is frequency dependent and sign reversal or zero-crossover effect at late decay times. A SQUID based TDEM system has been developed, characterized with prototype transmitter and field survey has been performed. Since the laboratory tests were done in a relatively noisy environment, the frequency dependence and the sign reversal effects have not been observed but the same have clearly been observed in the field survey. In the field survey, the decay transients surprisingly reached negative values in a terrain where a thin conducting layer is sandwiched between thick upper and thin lower resistive layers and the decay transients were highly dependent on base frequency of the transmitter waveforms. The major advantage of using the SQUIDs in TDEM applications is its extremely high sensitivity maintained at low frequencies. Therefore one can investigate the conductivity of the layers buried in deep below the surface of the ground. Nevertheless, the decay transients recorded with SQUID provided the information of the conducting layer along with an unexpected zero cross-over at later decay times. The data has been extensively analyzed and interpreted with the information available in the literature. At the end, as usual we suspected the sign reversal is due to instrumental drift, poor time synchronization between the transmitter controller and data acquisition system, remnant responses of the previous transmitter pulses, IP effect, poor offset removal, etc,. Finally, we found the genuine reason and it is hard to believe that the two peculiarities observed in TDEM survey with SQUID are due to the tertiary magnetic field (!) generated by the transmitter loop. In this article, we describe the possible reasons for these peculiarities based on the observations made by various researchers, understanding of the TDEM concepts, behavior of the instruments and various experiments performed by us.

## I. INTRODUCTION

Electromagnetic techniques are widely used in geophysical exploration applications and the response of these techniques is directly related to the surface and sub-surface properties of the earth, in particular electrical resistivity [1]. The recent introduction of SQUID (Superconducting Quantum Interference Device) magnetometer in different applications became popular due to its extremely high sensitivity for the tiny changes of magnetic field which is maintained even at very low frequencies [2-3].

The use of SQUIDs in TDEM (Time Domain Electro-Magnetic) geophysical exploration applications has potential advantage as compared to the conventional induction coil in addition to its sensitivity, wide bandwidth and wide dynamic range [4-7]. SQUIDs can measure the magnetic field directly whereas the induction coil measures the rate of change of magnetic field. In transient measurements, the voltage induced in the induction coil due to the secondary magnetic field (ground response) decays much faster and reaches the noise floor than the decay of the magnetic field itself. Therefore, the use of B - field sensor provides better target resolution and larger investigation depth. However, two peculiar effects such as frequency dependence and zero-crossover were observed with the SQUID based in-loop TDEM measurements. The response of the SQUID based TDEM signal recorded after the current passed through the transmitter loop is switched off depends strongly on the base frequency of the transmitter current waveform. The decay of the SQUID signal is much faster for higher base frequencies. Similarly, the response of the SQUID signal reaches a negative value at later decay times. These strange behaviors have been frequently observed in SQUID based transient measurement systems by various research groups and not in the systems operated with an induction coil sensor [5-8]. However, the above mentioned strange effects theoretically cannot occur in in-loop and coincident loop TDEM configurations [9-10]. The possible reason for these peculiar effects has been explained by G. Panaitov et al and H –J Krause et al [11-13]. They proposed a model for magnetometer based TDEM measurements that provide a possible explanation for these two effects. This model takes into account remnant responses induced in the ground by repetitive transmitter signals. According to them, in two layered ground with an upper resistive layer overlying the highly conducting basement, one can get much slower transient decay at intermediate time. Therefore, the positive and the non-negligible small negative current ring can persist simultaneously in the ground. The remnant non decayed persistent current rings disturb the recording of the original ground response. These disturbing responses will be even more for the B-field sensor particularly for SQUIDs due to its high sensitivity. They also discussed the procedures in order to optimize TDEM data collected with SQUID magnetometers based on the model proposed by them. Nevertheless, B R

Spies [14] agreed that the sign reversal occurring in transient decay at later times is genuine and disagreed with the mechanism proposed to explain the sign reversal and the procedure to optimize the TDEM data by G. Panaitov et al.

In this article, we describe the possible reasons for the above mentioned peculiar effects through the central loop TDEM measurements made with the SQUID sensor. The outline of this article is as follows. For the sake of completeness and the reader's convenience, first we describe the working principle of the TDEM with schematics. Then we describe the motivation of the experiment and how the traditional TDEM waveforms modified if the transmitter loop generates tertiary magnetic field and it couples to the receiver. Further, we describe the terrain location and its details where the TDEM central loop sounding measurements were performed. Subsequently, the detailed experimental data responsible for the so called peculiar effects have been presented. The analysis of the experimental data has also been provided with sufficient interpretation and conclusion.

## II. WORKING PRINCIPLE OF TDEM TECHNIQUE

The basic principle of TDEM is the induction of eddy currents in the ground with suitable excitation and subsequent measurement of its decay with a suitable sensor in time domain. Since the rate of decay of the eddy current strongly depends on the electrical conductivity of the target one can obtain information of the buried target. The schematic representation of the TDEM waveforms such as transmitter current in trapezoidal form to induce eddy currents in the earth and the response of the ground measured by the induction coil (which measures the rate of change of magnetic field) and SQUID (B-field sensor which directly measures the magnetic field) are shown in fig. 1. The primary magnetic field generated by the transmitter loop, the response of the ground in the form of secondary magnetic field and the net magnetic field measured by the magnetometer with respect to time are shown in fig. 1 (a), (b) and (c) respectively. Similarly, the primary magnetic field generated by the transmitter loop and the corresponding induced voltage in the induction coil sensor due to the primary magnetic field are shown in fig. 1 (d). The expected induced voltage in the induction coil due to the secondary magnetic field (target response) is shown in fig. 1(e). Since the target response alone cannot be measured by the induction coil (or magnetometer), the net voltage induced in the induction coil is shown in fig. 1(f). The advantages of using B-field sensor in TDEM measurements over an induction coil, selecting appropriate base frequency to remove power line noise by data processing, stacking data over a long period of time for averaging in order to suppress the background noise and final time window averaging are described in detail elsewhere [15].

## III. MOTIVATION OF THE EXPERIMENT AND MODIFIED TDEM WAVEFORMS WHEN TRANSMITTER LOOP GENERATES TERTIARY MAGNETIC FIELD

The SQUID based TDEM system for the use of geophysical application has been developed by integrating the SQUID system as a receiver with the other electronic instruments. The very first TDEM central loop sounding measurements in the field have been performed near Tummalapalle, Cuddapah (dt), Andhra Pradesh in the month of February 2019. This place was chosen based on prior information regarding the lithological structure of the region has been discussed in section 4. In this work, SQUID based TDEM central loop sounding measurements have been carried out with transmitter loops of different sizes and with different currents. These measurements were also performed with different base frequencies (2.5 Hz, 6.25 Hz, 12.5 Hz and 25 Hz). In these measurements, it is observed that the decay of the secondary magnetic field always reached negative values (zero crossover) at later decay times and the transients are strongly dependent on base frequency of the transmitter current. Similar zero crossover and frequency dependent TDEM transient responses have been observed with SQUID based central loop TEM measurements by Panaitov et al. [12] and Chwala et al. [7]. The data has been extensively analyzed and interpreted with the information available in the literature. Initially, we suspected the sign reversal is due to instrumental drift, poor time synchronization between the transmitter controller and data acquisition system, remnant responses of the previous transmitter pulses (as proposed by Panaitov et al. [12]), IP effect, poor offset removal, etc., [13]. The magnitude of the signal involved in the instrumental drift and offset variation over the time period of stacking are very low as compared to the magnitude of the signal reached to negative values at later decay times. The time synchronization between the transmitter controller and data acquisition system is excellent and the drift value between them is always better than ± 0.001 µs/hr. We also believe that there is a slim chance for the accumulation of the remnant eddy currents of the previous transmitter pulses. In view of the above reasons we suspect that there is some other reason for the zero crossover effects. We found the most probable reason for the zero crossover effects on the basis of experimental information available from the literatures [7, 12]. They found that this phenomenon is visible in a terrain with an upper resistive layer (where the transient is decaying fast) overlying the conducting layer. The terrain where we performed our TDEM experiments is also similar one and the reason for the two peculiarities observed in central loop TDEM measurements with SQUID is due to the tertiary magnetic field (!) generated by the transmitter loop. The generation of the tertiary magnetic field by the transmitter loop and coupling to the receiver such as magnetometer and induction coil are described with TEM waveform schematics.

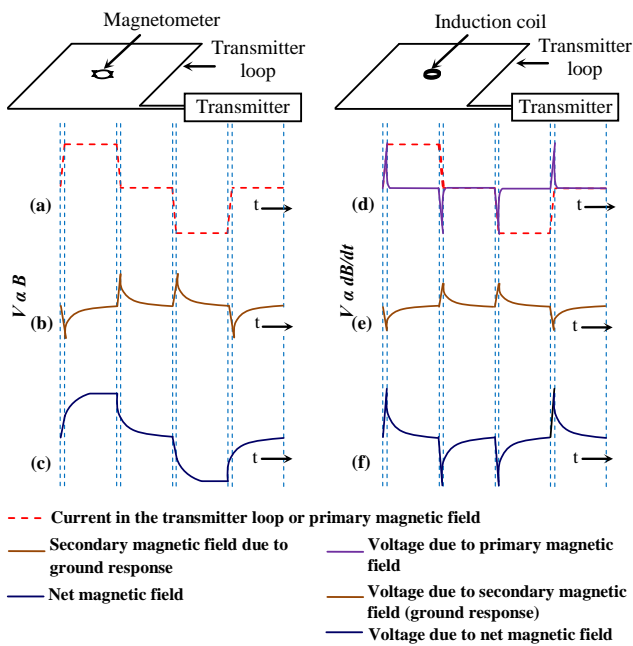

Fig. 1 Schematic sketches of the eddy current induction in the earth and the response of the magnetometer and an induction coil. (a) Current in the transmitter loop or primary magnetic field produced by the transmitter. (b) Secondary magnetic field due to the target. (c) Net magnetic field measured by the magnetometer. (d) Current in the transmitter loop and voltage induced in the induction coil due to primary magnetic field (e) voltage induced in the induction coil due to secondary magnetic field and (f) voltage induced in the induction coil due to net magnetic field.

The variation of TDEM waveforms or decay transients have been prepared when the transmitter loop partly acts as an induction coil and it generates tertiary magnetic field which is again coupled to the receiver located at the centre of the transmitter loop. The variations of decay transients with the use of SQUID magnetometer as well as an induction coil as a receiver are shown in fig. 2. The net magnetic field measured by the SQUID in regular TDEM central loop measurements is shown in fig. 2 (a). If the transmitter loop partly acts as an induction coil, the net voltage induced in the transmitter loop (fig. 2 (b)) and its associated current flows in the transmitter loop (in addition to the current pumped from the transmitter during ramp ON and ON time). This current produces additional magnetic field (fig. 1 (c)) which is again coupled to the magnetometer with respect to time. The additional magnetic field coupled to the SQUID magnetometer modifies the original decay throughout the transient and the decay may reaches to negative values at later decay times (fig. 2 (d)).

In a similar way, the net voltage induced in the induction coil in regular TDEM central loop measurements is shown in fig. 2 (e). At the same time, the voltage induced in the transmitter loop due to its behaviour as an induction coil and its associated current flow produces tertiary magnetic field and the same has been shown in fig. 1 (f). This additional tertiary magnetic field induced voltage (fig. (g)) which is again added to the induction coil and the total voltage induced in the induction coil modifies the entire original decay transients. The decay transients may also reaches negative values at later times.

The decay transients described here is only schematics to understand the signal coupling to the receivers such as magnetometer which produces voltage output proportional to magnetic field (**B**) and an induction coil produces output voltage proportional to d*B*/dt. The nature of the decay is mainly dependent on target nature (conductive or resistive) and its location (different depths) in addition to the resistance and inductance of the transmitter loop. Substantial amount of zero crossover effects have not been observed by the researchers and the survey personnel involved in ground based TDEM measurements with an induction coil as a receiver. Nevertheless, the negative decay transients have been observed in TEM survey with coincident loop configuration and these transients have been modelled with polarisable half planes [16]. John Bishop et al. described several in-loop TEM survey examples with negative decay transients in a highly resistive terrain where there is little or no polarisable material present. He also suspected that there is something lying within the transmitter and listed many including "*the possibility of the reverse current flow in the transmitter during the OFF time or the receiver and a previously unrecognised physical effect*" [17]. Sean E. Walker also observed such negative decay transients even at very earlier time in helicopter TEM survey [18]. Similar types of negative transients have been observed worldwide with several groups in TEM survey performed with central loop or coincident loop.

In view of this above information, we had repeated the central loop TDEM measurements in the same place in order to measure the current flow through the transmitter loop and measure the decay of the secondary magnetic field with SQUID at the centre of the loop simultaneously. Last year, we were not able to conduct measurements due to COVID 19 pandemic and this year we performed the experiments with the transmitter loop sizes of 100 m x 100 m and 400 m x 400 m. The experimental details and analysis of the experimental results are presented in following sections.

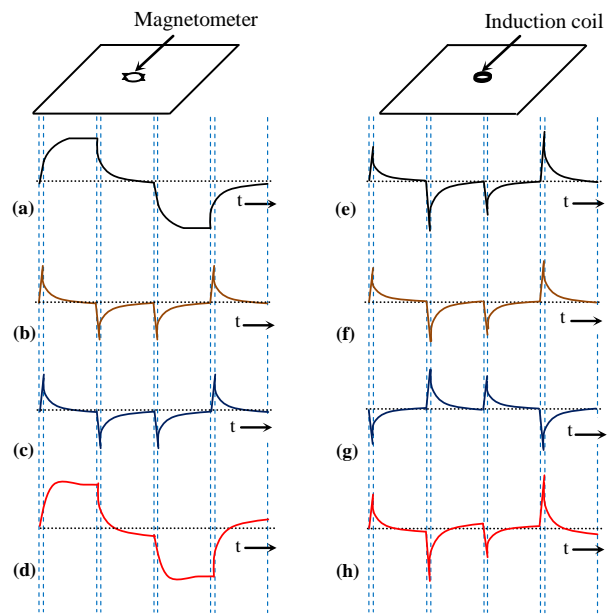

Fig. 2 The variations of decay transient waveforms with the use of magnetometer and an induction coil as a receiver when transmitter loop partly acts as an induction coil.
(a) Net magnetic field measured by the SQUID without additional signal coupling from the transmitter loop. (b) Voltage induced in the transmitter loop due to the behaviour of the induction coil (c) and its associated current produces tertiary magnetic field and (d) total magnetic field measured by the magnetometer (summation of (a) and (c)).
(e) Net magnetic field measured by the induction coil without additional signal coupling from the transmitter loop. (f) Voltage induced in the transmitter loop due to the behaviour of the induction coil (g) and its associated current produces tertiary magnetic field which is again coupled (g) as a voltage to the induction coil and (h) total voltage induced in the induction coil (summation of (e) and (g)).

IV. EXPERIMENTAL SITE AND TERRAIN DETAILS

The SQUID based TDEM central loop sounding measurements have been performed near Tummalapalle Uranium mine. Tummalapalle uranium mine is located in the Cuddapah district of Andhra Pradesh, in the southern part of India. Geologically, Tummalapalle is in the south western part of the Cuddapah basin. Dolostone hosted uranium mineralization occurs in the Vempalle Formation of Cuddapah basin intermittently over a strike length of 160 km from Reddypalle in NW to Maddimadugu in the SE [19].

Lithologically, the Vempalle formation consists of massive limestone, purple shale, intra-formational conglomerate, dolostone (uraniferous), shale and cherty limestone. The impersistent conglomerate and grey shale band occurring immediately below and above the mineralized rock respectively, serve as the marker horizons [20]. The stratigraphy of the terrain where the SQUID based TDEM central loop sounding measurements has been performed is shown in fig. 3. Here, the cherty Limestone covered in the upper layer is thick (~ 635 m) and highly resistive. The shale layer located just below the Limestone is thin (16 m) and conductive. As far as TDEM measurements are concerned, the layers of the earth have been considered as upper conductive weathered, resistive overburden (Cherty Limestone), thin conducting layer (shale) and the remaining layers are resistive. The stratigraphy of the terrain has been obtained from the borehole data collected by the geologists.

V. EXPERIMENTAL DETAILS

The TDEM system comprises of a transmitter (ZT-30 Zonge International) to drive a current in the form of trapezoidal pulses through the transmitter loop with the help of external batteries, transmitter controller (SMARTem24 - EMIT) to set and control the waveform parameters in the transmitter, transmitter loop and fast data acquisition system (SMARTem24 - EMIT) to record the response of the sensor. The transmitter can supply a maximum current of 30 A. The data acquisition system can display the recorded data in the form of raw, stacked, decay profile and station profile by using built-in software and store it for further processing. Here, the cryostat containing the SQUID probe has been partially buried in the ground in order to arrest the wind generated mechanical vibrations. The schematic diagram of the experimental setup and the photograph of the experimental site along with the coordinates of the 400 m square transmitter loop are shown in fig. 4.

A transmitter loop in the form of square in which a resistor with a value of 0.1 ohm is connected in series has been laid in

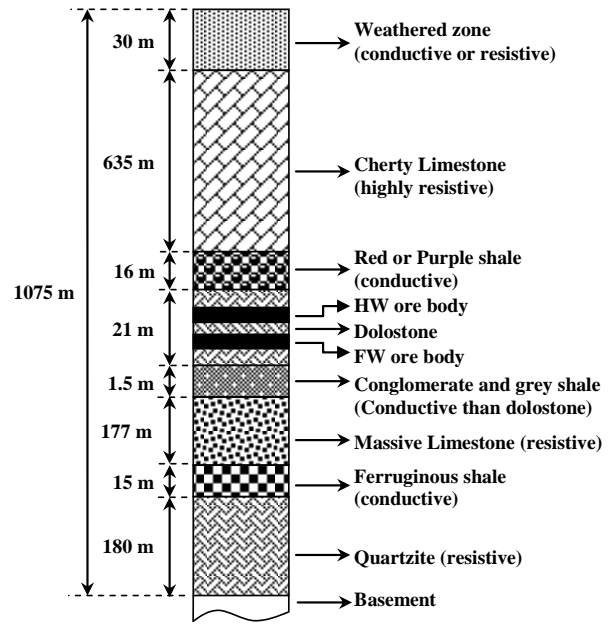

Fig. 3 The stratigraphy of the terrain

the field. The liquid helium cryostat with SQUID which measures the vertical component of the magnetic field has been located at the centre of the transmitter loop. A wire in the form of twisted pair has been used to measure the voltage drop across the 0.1 ohm resistor. The twisted pair from 0.1 ohm

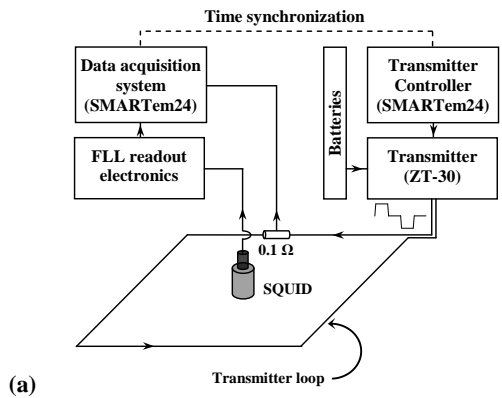
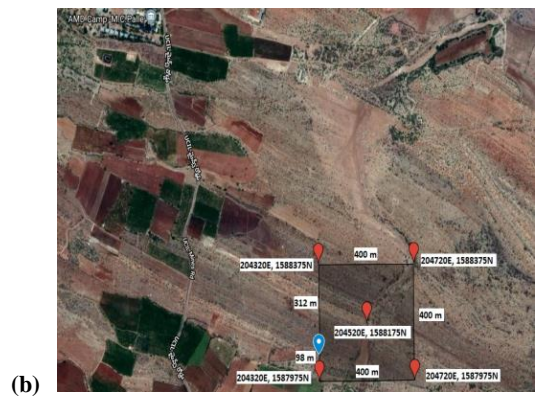

Fig. 4 (a) Schematic diagram of the experimental set up. (b) The photograph of the experimental site along with the coordinates of the 400 m square transmitter loop.

resistor has been brought to the centre of the transmitter loop in order to measure the SQUID output and the voltage output across 0.1 ohm by keeping the data acquisition system at the same place. The SQUID measures magnetic field in units of pT based on the flux to field transfer coefficient of the SQUID (3 nT/$\Phi_0$) and gain of the FLL (V/$\Phi_0$) and the voltage across the 0.1 ohm measures in units of µV.

## VI. EXPERIMENTAL RESULTS AND DISCUSSION

Initially, transmitter loop in the form square with side length of 100 m has been laid. Two copper wires (copper wires generally used for commercial purposes) each with cross sectional area of 6 square mm are connected in parallel in order to reduce the total resistance of the loop and its measured value is around 1.1 ohms including 0.1 ohm resistor connected in series with the transmitter loop. The base frequency of the transmitter current, sampling rate, duty cycle and number of stacks are kept at 1.25 Hz, 24000/s, 50% and 2048 respectively. The same experiments have been repeated with different base frequencies (2.5 Hz, 6.25 Hz, 12.5 Hz, 25 Hz and 37.5 Hz) and the other parameters are maintained at the same values.

### ANALYSIS OF ON TIME DATA

It is well known that the data recorded in TEM measurements just after the transmitter current is OFF is always used for processing to obtain the geological structure of the terrain. In this article, we analysed both ON time transients as well as OFF time transients to understand the frequency dependent and zero crossover effects. The voltage transients measured across 0.1 ohm resistor connected in series with the transmitter loop and the magnetic field transients measured by the SQUID during the transmitter current is ON for different base frequencies are shown in fig. 5 (a) and (b) respectively. The insets are the zoomed views during the transmitter current is ON indicating that the voltage across 0.1 ohm resistor and the magnetic field measured by the SQUID are base frequency dependant. The net voltage recorded across 0.1 ohm resister comprises of voltage induced due to the changes of the primary magnetic field and voltage induced due to the changes of the secondary magnetic field (terrain response) in addition to the voltage drop due to the applied transmitter current. The voltage measured across 0.1 ohm resistor due to the transmitter current is proportional to the current flow in the transmitter loop and the corresponding magnetic field measured by the SQUID are simultaneous process with respect to time. Therefore, this can be simply ignored for the present discussion. The induced voltage due to the changes of the primary magnetic field is dependent on the resistance and inductance of the transmitter loop. Similarly, the voltage induced due to secondary magnetic field which is terrain dependant is also added to the transmitter loop and the net voltage is measured across 0.1 ohm resistor. The net voltage drop measured across 0.1 ohm resistor for different base frequencies are shown in fig. 5 (a). With careful observation of the data in fig. 5 (a), it has been seen that there was a crossover between the transients with low base frequency and high base frequency. Since the changes of the voltage values on both sides of the crossover are extremely small, it has been schematically shown in fig. 6 and the time at which the crossover occurred between the transients have been extracted from the data. The crossover time between the transients with the base frequencies of 1.25 Hz and 2.5 Hz, 2.5 Hz and 6.25 Hz and 6.25 Hz and 12.5 Hz are 3.95 ms, 1.88 ms and 0.91 ms respectively. The time to crossover occurred at different times for different base frequencies clearly indicates the generation of the secondary magnetic field and its associated induced voltage in the transmitter loop due to the

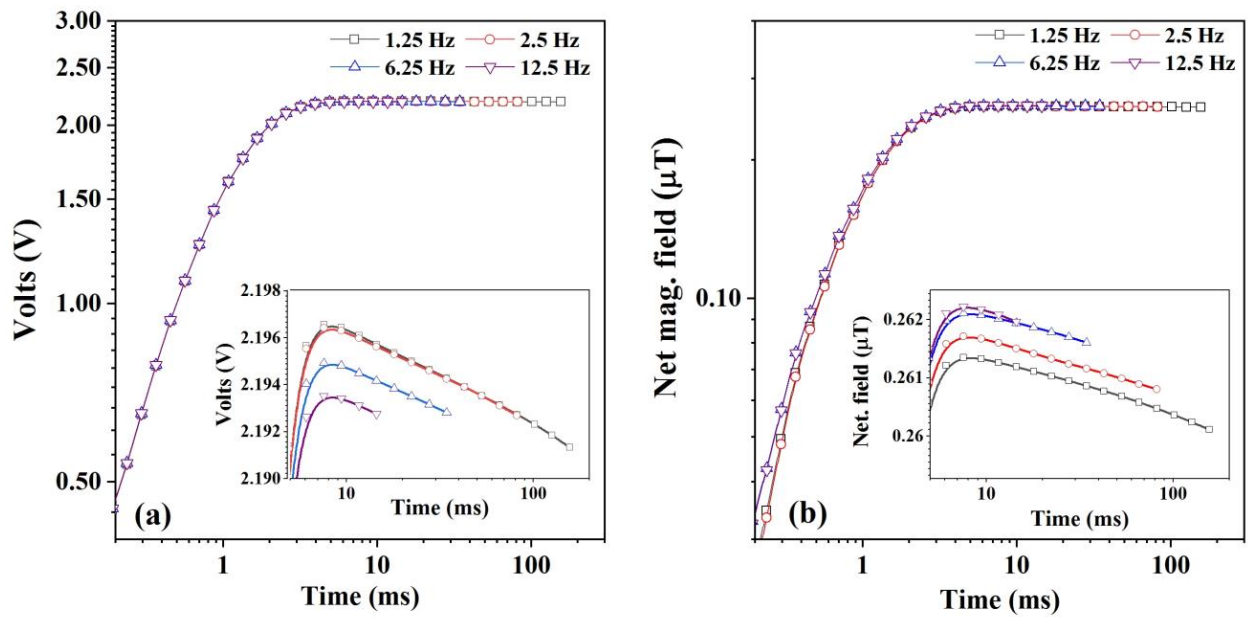

Fig. 5 (a) Voltage measured across 0.1 ohm resistor connected in series with the transmitter loop and (b) magnetic field measured by the SQUID located at the centre of the transmitter loop during the transmitter current is ON for different base frequencies. The insets are the zoomed views during transmitter current is ON indicating that the voltage across 0.1 ohm resistor and the magnetic field measured by the SQUID are base frequency dependant presence of conducting layer in the terrain. Nevertheless, the dominant voltage induced in the transmitter loop is due to the changes of the primary magnetic field which is proportional to the applied current in the transmitter loop. From fig. 5 (a) it is clear that the magnitude of the induced voltage is proportional to the net magnetic field and is low for higher base frequencies and vice versa. For low base frequency sufficient time is available to attain maximum current and the corresponding voltage induced in the transmitter loop is high. For base frequencies above 12.5 Hz, the current in the transmitter loop not even reached to its maximum value (say 25 Hz and 37.5 Hz).

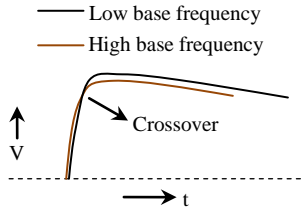

Fig. 6 Schematic view of the crossover between the transients with different base frequency

Similarly, it is well known that there will not be any crossover for the SQUID located at the centre of the transmitter loop with different magnitude of the applied transmitter current for the terrain which is either conductive or resistive. The SQUID measures the net magnetic field with respect to time that is the primary magnetic field and the secondary magnetic field produced due to the terrain. The net magnetic field is frequency dependant and is expected to be similar to the voltage transient measured across 0.1 ohm resistor. But actually it is reverse that the amplitude of the SQUID signal is maximum for high base frequency and the same has been shown in fig. 5 (b). Therefore, it is suspected that some additional signal added to the SQUID from some other source. *That is from the transmitter loop.* The additional voltage induced in the transmitter loop due to the inductive coupling of the primary magnetic field and voltage induced due to the terrain (net voltage is proportional to the net magnetic field) and its associated current flow in the transmitter loop produces magnetic field which is additionally coupled to the SQUID with respect to time. The transmitter loop is performing partly as an induction coil sensor in which the induced voltage ($V(t) \alpha (\partial B/\partial t)$) and its associated current flow ($I(t) = V(t)/R$) produces magnetic field. Here, R is the resistance of the transmitter loop. Similarly, the voltage induced in the transmitter loop after the transmitter current off is also coupled to the SQUID and its effect is severe due to the abrupt switch off of the transmitter current.

This data has been further analysed by normalizing the peak values recorded across 0.1 ohm resistor and magnetic field recorded by the SQUID for different base frequencies and the same have been shown in fig. 7. Here, one can observe that the rate of increase of the magnetic field measured by the SQUID is high as compared to the voltage measured across 0.1 ohm resistor. In addition to this, the time to peak for the SQUID is always less than the voltage measured across 0.1 ohm resistor and surprisingly the net magnetic field measured by the SQUID decays much faster than the voltage measured by the 0.1 ohm resistor. This clearly indicates that the SQUID measures some extra magnetic field in addition to the primary magnetic field produced by the current applied to the transmitter and secondary magnetic field induced by the terrain. This extra magnetic field is from the transmitter loop due to the current induced in the transmitter loop. The schematic representation of the TDEM waveforms such as the current flow through transmitter loop, voltage induced in the transmitter loop, the expected net magnetic field measured by the SQUID including the terrain response and the actual magnetic field measured by the SQUID during the transmitter current is ON are shown in fig. 8 (a), (b), (c) and (d) respectively. The time to peak values obtained from the transients recorded across 0.1 ohm resistor and SQUID after normalizing to their peak values for different base frequencies of the transmitter current are tabulated in table I.

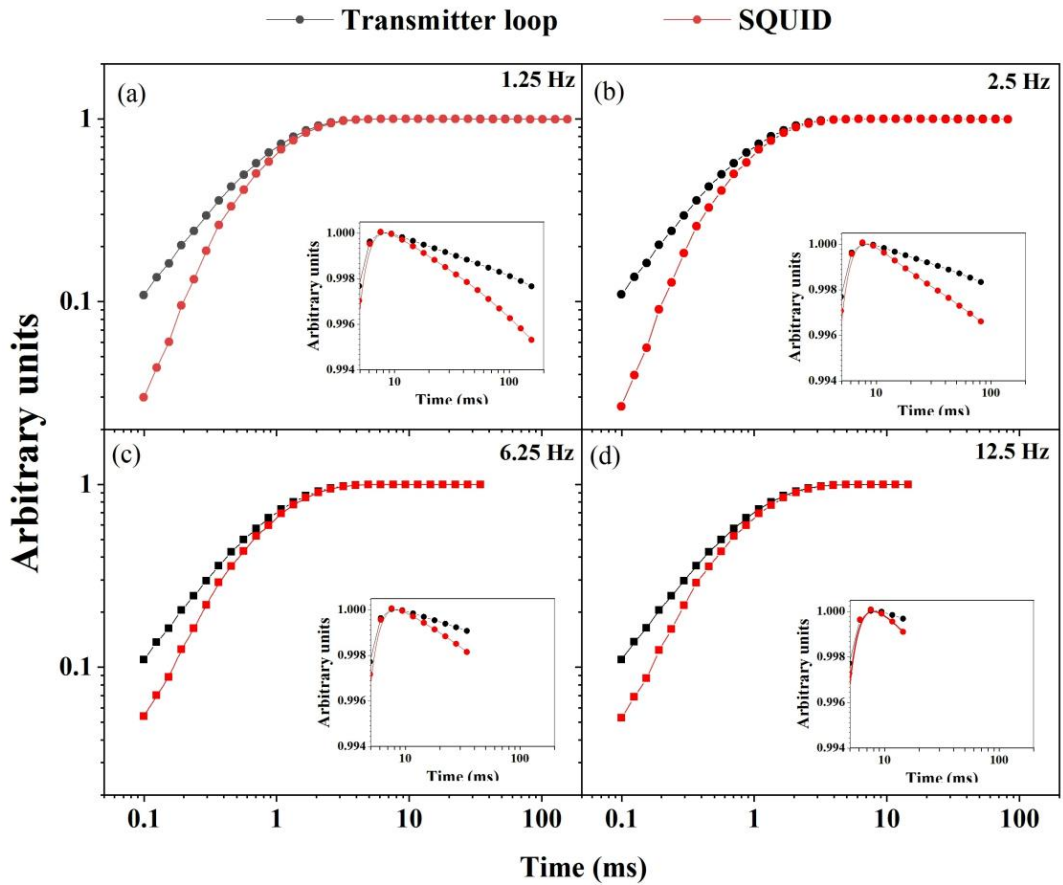

Fig. 7 The recorded voltage across 0.1 ohm resistor and SQUID output at the centre of the transmitter loop are normalized to their peak values for different base frequency of the transmitter current (a) 1.25 Hz, (b) 2.5 Hz, (c) 6.25 Hz and 12.5 Hz.

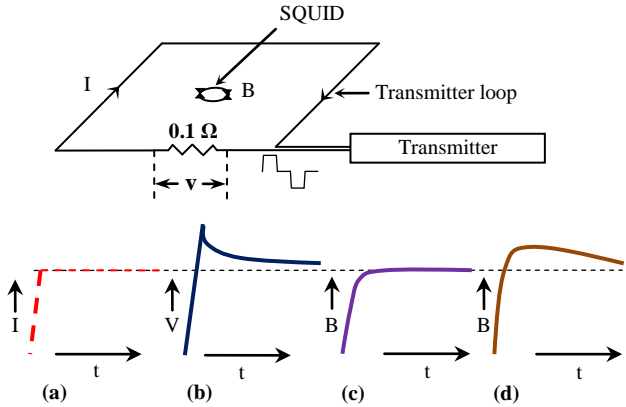

Fig. 8 Schematic view of the TDEM waveforms during the transmitter current is ON. (a) Current flow through the transmitter loop (b) resultant voltage across the transmitter loop (c) expected magnetic field measured by SQUID including the terrain response and (d) the resultant magnetic field measured by the SQUID magnetometer.

Table I: Time to peak values obtained from the transients recorded across 0.1 ohm resistor and SQUID after normalizing to their peak values for different base frequencies of the transmitter current.

| Frequency (Hz) | Time to peak (ms) | |
|---|---|---|
| | Across 0.1 ohm | SQUID |
| 1.25 | 8.35138 | 7.85736 |
| 2.50 | 8.36598 | 8.10044 |
| 6.25 | 8.39512 | 8.20336 |
| 12.5 | 8.42707 | 7.94988 |

The same experiments have been repeated with just 16 stacks and the ON time transients recorded across 0.1 ohm resistor and SQUID for different base frequencies are shown in fig. 9 (a) and (b) respectively. Here also we clearly observed the inverse frequency dependence of the transients between the voltage recorded across 0.1 ohm resistor and SQUID. In addition to this, one can see that the secondary magnetic field recorded with SQUID is decaying much faster from its maximum value than the voltage recorded across 0.1 ohm resistor and the same has been observed in fig. 7 with more averaging. It is inferred that the fast decay is due to the ground response (second resistive layer) reflected in the transmitter loop which is again coupled to the SQUID as a tertiary magnetic field. The peak voltage recorded across 0.1 ohm resistor with base frequency of 1.25 Hz is slightly higher

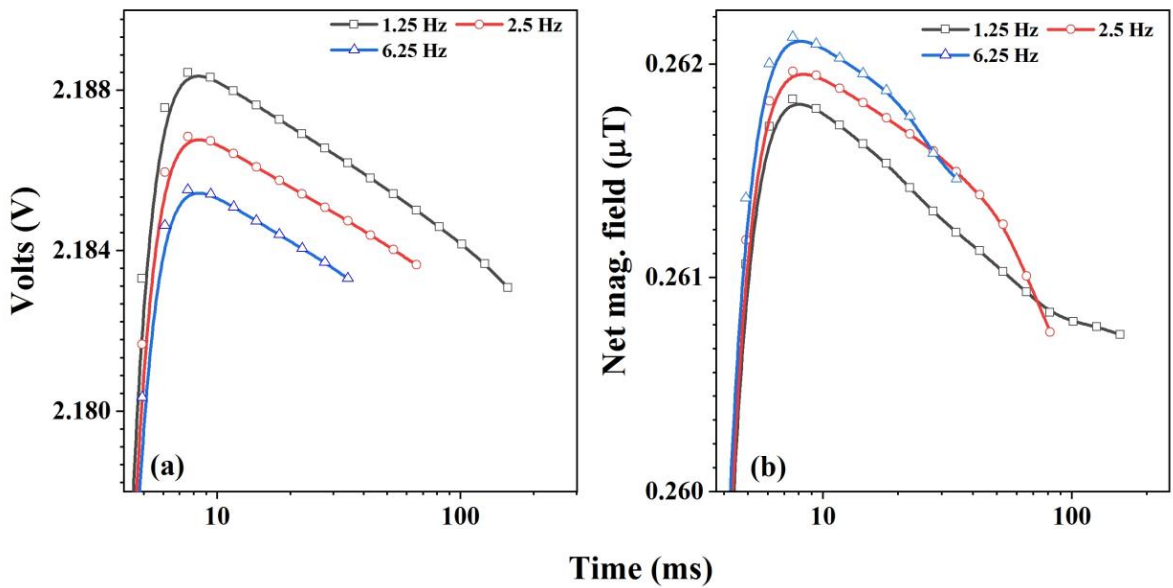

Fig. 9 (a) Voltage measured across 0.1 ohm resistor and (b) magnetic field measured by the SQUID with just 16 stacks during the transmitter current is ON for different base frequencies.

than 2.188 V where as in fig. 5 it is slightly higher than 2.196 V which is due to higher number of stacks of 2048.

From the above analysis, it is very clear that a strong voltage induced in the transmitter loop is due to the combination of the primary and secondary magnetic field. This voltage and its associated current produce tertiary magnetic field which is further coupled to the sensitive receiver SQUID. Moreover this net voltage induced in the transmitter loop is not decaying to zero and is transferring to the next OFF cycle. The data analysis during the OFF time of the transmitter current has been discussed in the next section.

## ANALYSIS OF OFF TIME DATA

The OFF time decay transients recorded by the SQUID at the centre of the transmitter loop and the voltage drop measured across 0.1 ohm resistor connected in series with the transmitter loop at different base frequencies are shown in fig. 10 (a) and (c) respectively and the corresponding zoomed views in the region where the crossover effects were occurred are shown in fig. 10 (b) and (d). Since the data after the zero crossover has negative values, we have simply taken the screen shots from the SMARTem24 data acquisition system and presented for discussion. Nevertheless, the essential and sufficient information is explicitly visible in the decay data. The decay transients recorded two years ago and the data recorded this year in the same place are the same. The ramp off time for this 100 m x 100 m square transmitter loop and the applied current of about 22 A is about 300 μs. For comparison of the SQUID data and the voltage data recorded across 0.1 ohm resistor, the data has not been processed with the ramp off time. From fig. 10 (a), it is observed that the secondary magnetic field recorded with SQUID decays rapidly up to 0.695 ms after the ramp off time and slows down from 0.695 ms to 1.071 ms. After 1.071 ms, the secondary magnetic field abruptly changed to a large negative value and thereafter, reached zero over a long period of time. One can predict the rapid decay up to 0.695 ms after the ramp off corresponding to the upper resistive layer and the slow decay between 0.695 ms to 1.071 ms corresponding to the thin conducting layer. After 1.071 ms, the rapid decay corresponds to the second resistive layer. The SQUID signal is very clear from 1.071 ms to 155 ms (for base frequency of 1.25 Hz) and this can be clearly seen in the zoomed view in fig. 10 (b).

Similarly, from fig. 10 (c) it is observed that voltage induced in the transmitter loop (measured across 0.1 ohm resistor) is in the reverse direction just after the ramp off (refer fig. 1 (d)). Here, one can also observe the clear slope change between 0.695 ms and 1.071 ms which indicates the presence of conductive layer and the sensing capability of the transmitter loop (acts as an induction coil). After 1.071 ms, it is expected that the voltage has to decay towards zero. Instead, the voltage again jumped to positive values at the exact time when the SQUID output moved from positive to negative values. There may be a slight delay between the two transients in this region but that is not visible in the data which has been window averaged. The reason for the voltage jump from negative value to positive value is due to the large voltage change due to the second resistive layer and the non-zero decay of the voltage induced in the transmitter loop due to the previous positive ON cycle. The schematic representation of the TDEM waveforms such as the current flow through transmitter loop, voltage induced in the transmitter loop and the non-decayed voltage in the transmitter loop during ON time which is extending to the OFF time are shown in fig. 11 (a), (b) and (c) respectively. Similarly, the expected net magnetic field measured by the SQUID including the terrain response and the actual magnetic field measured by the SQUID during the transmitter current is OFF are shown in fig. 11 (d) and (e) respectively.

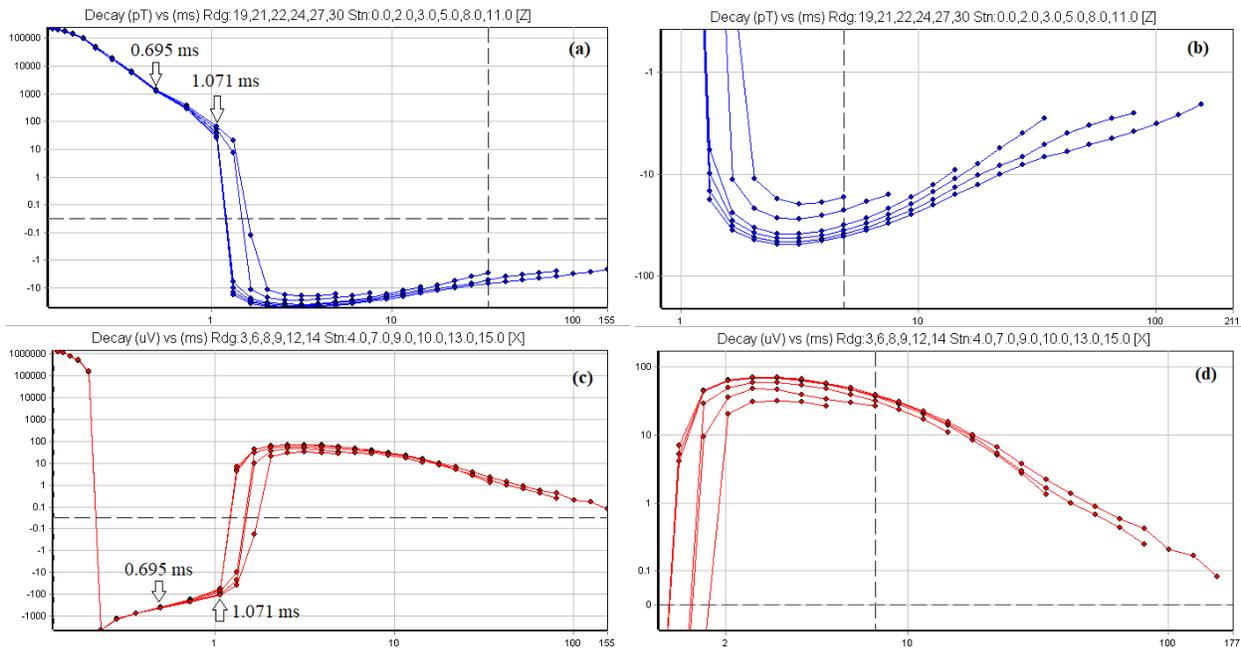

Fig. 10 Decay transients recorded by the SQUID at the centre of the transmitter loop and the voltage drop measured across 0.1 ohm resistor connected in series with the transmitter loop at different base frequencies are shown in (a) and (c) respectively and the corresponding zoomed view are shown in fig (b) and (d).

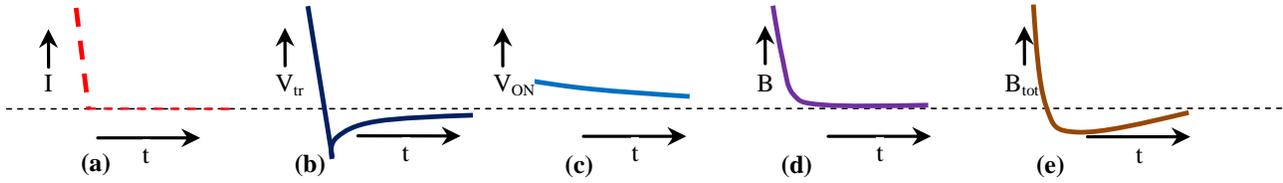

Fig. 11 Schematic view of the TDEM waveforms during the transmitter current is OFF. (a) Current flow through the transmitter loop (I), (b) resultant voltage across the transmitter loop ($V_{tr}$), (c) non-decayed voltage in the transmitter loop during ON ($V_{ON}$) time which is extending to the OFF time, (d) expected magnetic field measured by SQUID (B) and (e) the resultant magnetic field measured by the SQUID magnetometer ($B_{tot}$).

In addition to this, one can simply visualize and compare the quality of the data recorded with SQUID which is much clearer than the data recorded across 0.1 ohm resistor even up to the last time window (155 ms corresponds to the base frequency of 1.25 Hz). This shows that the magnetic field coupled to the SQUID which is generated by the transmitter loop due to the reverse current flow is much stronger than the secondary magnetic field directly coupled to the SQUID due to the response of the ground. The magnitude of the signal for the lower base frequency is high indicating that the applied excitation current is more as compared to the excitation current available with higher base frequency. Hence, the effect of frequency dependence in TDEM measurements is simply due to the magnitude of the excitation current which is inversely proportional to base frequency.

For better clarity, the SQUID transients and its associated voltage transients recorded across 0.1 ohm resistor are plotted separately for different base frequencies and the same have been shown in fig. 12. The crossovers between the SQUID transients and the voltage transients clearly show that the dominant signal coupled to the SQUID is due to the reverse current flow through the transmitter loop. The discussion on the flow of the reverse current in the transmitter circuit including transmitter loop, IGBTs with built-in diodes, synchronization of timing pulses, etc., are not included in this article due to the size of the article and ethics followed by the use of advanced technologically important commercial equipments.

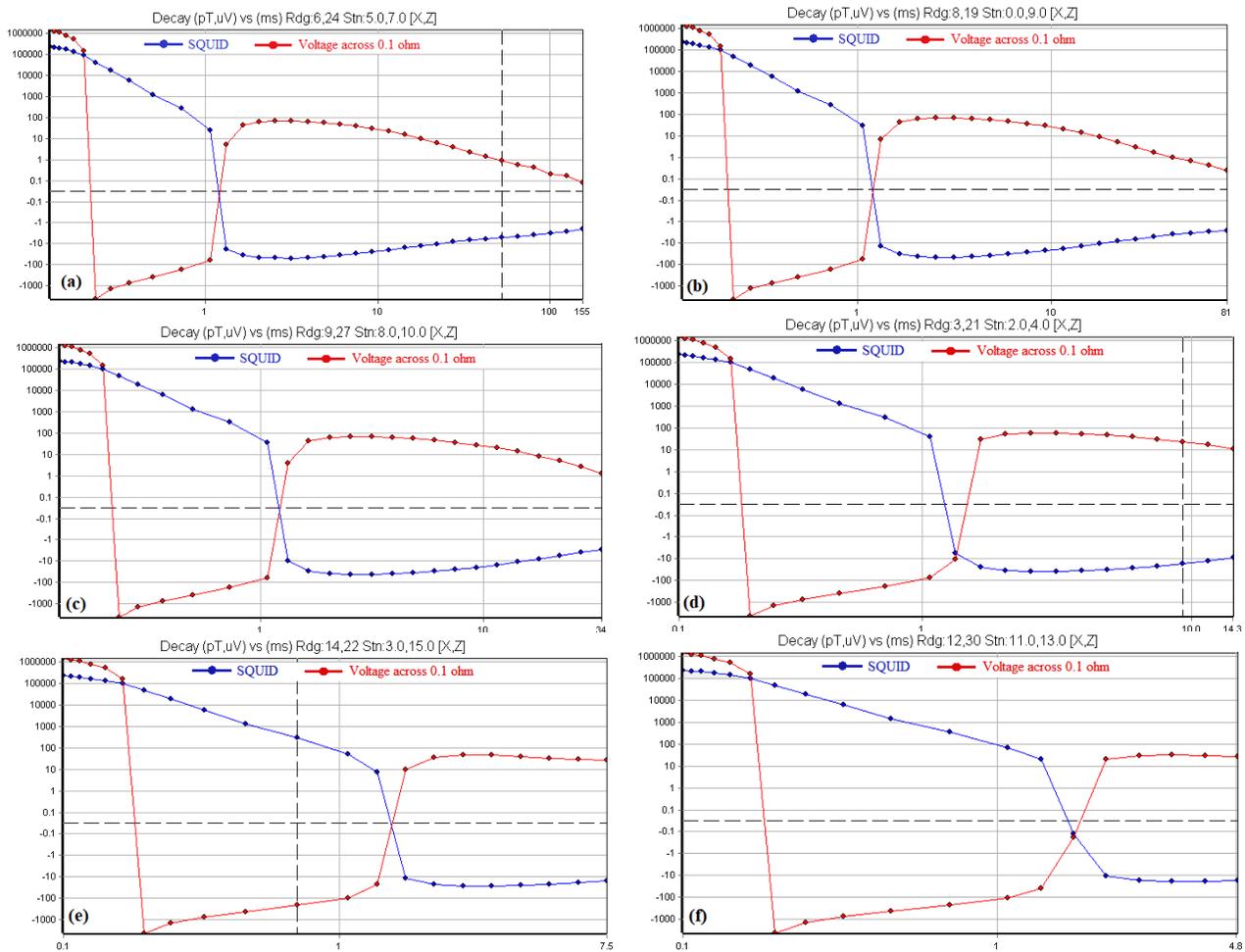

Fig .12 SQUID transients and its associated voltage transients recorded across 0.1 ohm resistor for different base frequencies (a) 1.25 Hz, (b) 2.5 Hz, (c) 6.25 Hz, (d) 12.5 Hz, (e) 25 Hz and (f) 37.5 Hz.

## VII. TDEM Measurements with Larger Transmitter Loop

Experiments have been further repeated by increasing transmitter loop size of 400 m x 400 m and with different base frequencies. In this experiment, we used part of the transmitter loop as a resistor and the voltage drop has been measured with time. The schematic view of the transmitter loop and the voltage measured across part of the transmitter loop are shown in fig. 13. The part of the transmitter loop whose length is about 11 m is used as a resistor. The dc resistance of the wire is about 27 mΩ which is independently measured in the laboratory by four probe method. Here, the applied current to the transmitter loop is around 26 A and the equivalent central field is around 78 nT. The SQUID transients and their associated voltage transients recorded across 27 mΩ resistor connected in series with transmitter loop of 400 m x 400 m for different base frequencies (a) 1.25 Hz, (b) 2.5 Hz, (c) 6.25 Hz, (d) 12.5 Hz, (e) 25 Hz and (f) 37.5 Hz are shown in fig. 14.

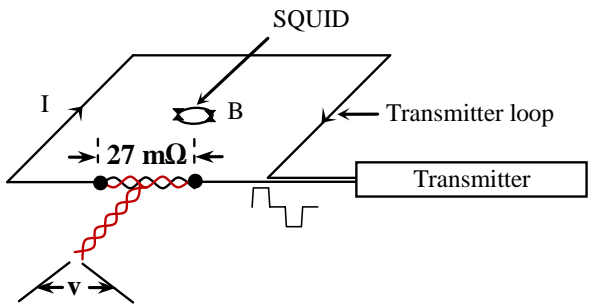

Fig 13 Schematic view of the transmitter loop and the voltage measured across part of the transmitter loop

Since the maximum investigation depth and the target resolution are proportional to transmitter moment, the slow decay of the secondary magnetic field which corresponds to the conducting layer has been clearly observed at a larger decay times as compared to the measurements performed with transmitter loop size of 100 m x 100 m. The signal coupled to the SQUID from the ground is also much better as compared to that coupled from the transmitter loop. But, still substantial amount of magnetic signal is coupled to the SQUID from transmitter loop which is deteriorating the SQUID performance. Here also the crossovers between the transients have been clearly observed. At the same time, transients with some base frequencies (say 12.5 Hz and 25 Hz), recorded across the transmitter loop ($V \alpha\ dB/dt$) decay much faster than the decay of the magnetic field measured by the SQUID ($V \alpha\ B$). This is also the indication that the use of transmitter loop with larger size reduces the signal coupling due to the reverse current induced in the transmitter loop.

## VIII. Additional Discussion and Summary

The important points have been summarised according to the experimental observations and analysis of the experimental data. The summary has been organized into two parts. The first part is about the observation of frequency dependence and zero crossover effects in ground based central loop TDEM measurements with SQUID and the same is not observed with an induction coil. The second part is our interpretation made on the basis of the research work done by others. Many survey groups observed negative transients at very early decay time itself or at later decays times with in-loop and coincident loop TEM systems including helicopter based system.

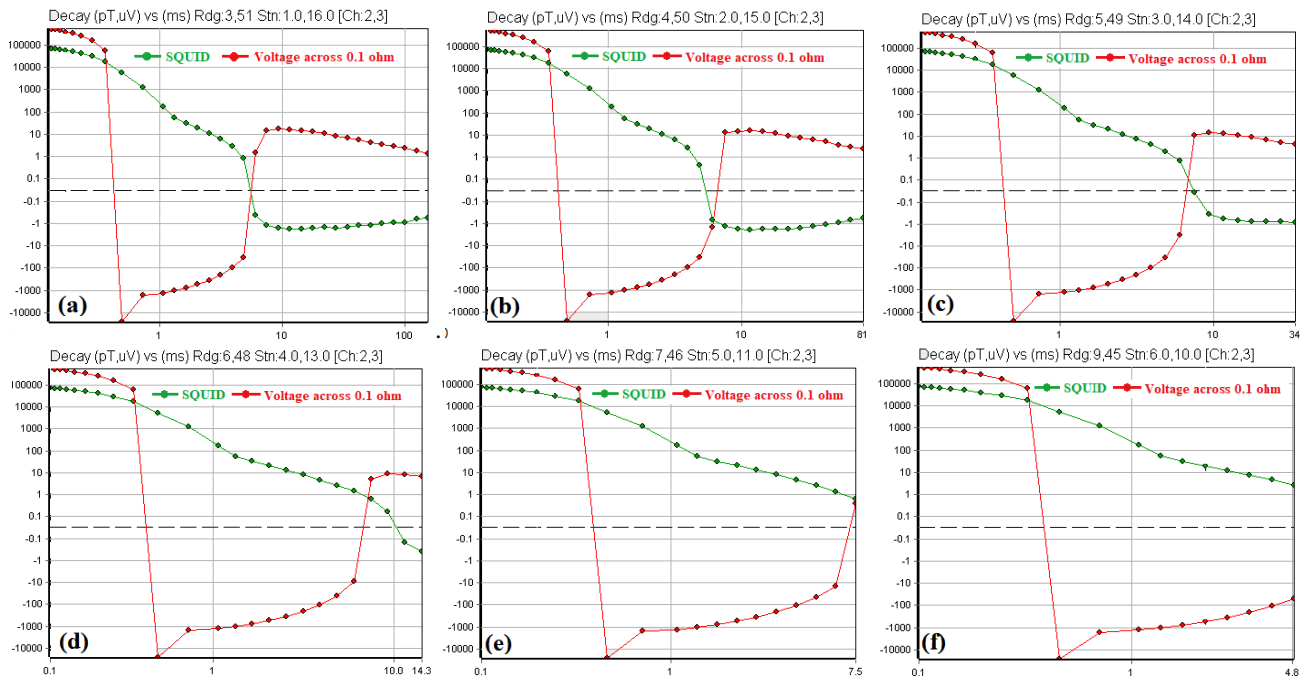

Fig .14 SQUID transients and its associated voltage transients recorded across 27 mΩ resistor connected in series with transmitter loop of 400 m x 400 m for different base frequencies (a) 1.25 Hz, (b) 2.5 Hz, (c) 6.25 Hz, (d) 12.5 Hz, (e) 25 Hz and (f) 37.5 Hz.

The main reason for the observation of the frequency dependence and zero crossover effects with SQUID is only due to its extreme sensitivity maintained even at very low frequency. In general, the resistance value of the wire or cable used as a transmitter loop is low in order to pump more excitation current. Moreover, the turn ON time and turn OFF time of the transmitter current is different. Slow turn ON may induce less inductive current and it takes longer time to decay if the resistance of the transmitter loop is low. Therefore, this decay of the inductive current carries over to the next immediate OFF time. Similarly, during OFF time the induced inductive current and its decay are fast for the given resistance of the transmitter loop. This induced inductive current and its decay is sensed by the sensitive sensor like SQUID. This is what we have observed as zero crossover effects in the present work. Similarly, the excitation current may not reach its steady value for the higher base frequency and therefore, its associated inductive response including target response is also low. The variation of the excitation current amplitudes is slightly larger for higher base frequencies (say between 12.5 Hz and 25 Hz) as compared to the low frequencies (between 1.25 Hz to 2.5 Hz). The target response for such a slight variation of excitation currents could be sensed by the SQUID due to its sensitivity. Therefore, the frequency dependence is always observed with SQUID based TEM measurements irrespective of the nature of the targets. Similarly, the induction coil sensor is generally fabricated with a larger number of turns using thin wires and its overall dc resistance is high. Therefore, induction coil sensors are insensitive for the smaller inductive currents coupled either from the transmitter loop or from the ground.

At the same time, in some TEM systems where the inductive coupling is stronger between the transmitter loop and the receiver like helicopter based TEM systems and coincident loop TEM systems, the reverse current flow in the transmitter loop provides strong signal coupling to the receiver. Therefore, the negative transients occur at later time in some cases or even at very early times if the terrain is highly resistive.

## IX. CONCLUSION

The frequency dependence and the sign reversal effects have been observed in our first field survey with SQUID based TDEM central loop sounding measurements. The field is located far from the laboratory where the terrain consists of a thin conducting layer sandwiched between thick upper and thin lower resistive layers. The reasons for these effects have been speculated through analysis of the transient data recorded with SQUID and information available from the literature including TEM systems with an induction coil sensor. The speculation is the flow of reverse current in the transmitter loop which is further coupled to the sensitive magnetic field sensor SQUID which leads to zero crossover effects. The TDEM measurements have been repeated in the same location and this time a small resistor (0.1 ohm) has been connected in series with the transmitter loop to measure the inductive current flow during ON time as well as OFF time. The experimental results and analyses provide sufficient evidence to prove that the peculiarities such as the frequency dependence and the sign reversal effects in SQUID based TDEM measurements are real. Further work is essential to eliminate or at least to minimize the reverse current flow in the transmitter loop in order to utilize the entire capability of the SQUID sensor in geophysical measurements.

ACKNOWLEDGEMENTS

The authors would like to thank Shri R. Baskaran, Dr N.V. Chandra Shekar for their continuous support. The authors would like to extend special thanks to Dr. A. K. Bhadurai, Director, IGCAR and Dr. D.K. Sinha, Director, AMD for their continuous support and encouragement in this work. The authors would also thank Mr B.V. L Kumar, senior geophysicist, AMD for his guidance in the field. The authors would also thank to Dr K. Gireesan and Shri C.M. Kumar for providing liquid helium in time to perform these measurements outside the laboratory. The authors also thank Mr R. Ramkumar and Mr E. Goutham for their help and support in laying transmitter loops in a rough terrain